\documentclass[final,authoryear,1p,twocolumn]{elsarticle}
 \usepackage{graphicx,subfig}
\usepackage{amssymb, amsmath}

\journal{Journal of Theoretical Biology}

\begin{document}

\begin{frontmatter}

\title{Punishment can promote defection in group-structured populations}

\author[bath]{Simon T. Powers\corref{cor1}\fnref{fn1}}
\ead{Simon.Powers@unil.ch}

\author[bath]{Daniel J. Taylor}
\ead{D.J.Taylor@bath.ac.uk}

\author[bath]{Joanna J. Bryson}
\ead{J.J.Bryson@bath.ac.uk}

\cortext[cor1]{Corresponding author}
\fntext[fn1]{Present address: Department of Ecology \& Evolution, Biophore, University of Lausanne, CH-1015 Lausanne, Switzerland} 
\address[bath]{Department of Computer Science, University of Bath, Bath BA2 7AY, UK}

\begin{abstract}
  Pro-social punishment, whereby cooperators punish defectors, is
  often suggested as a mechanism that maintains cooperation in large
  human groups. Importantly, models that support this idea have to
  date only allowed defectors to be the target of punishment. However,
  recent empirical work has demonstrated the existence of anti-social
  punishment in public goods games. That is, individuals that defect
  have been found to also punish cooperators. Some recent theoretical
  studies have found that such anti-social punishment can prevent the
  evolution of pro-social punishment and cooperation. However, the
  evolution of anti-social punishment in group-structured populations
  has not been formally addressed. Previous work has informally argued
  that group-structure must favour pro-social punishment. Here we
  formally investigate how two demographic factors, group size and
  dispersal frequency, affect selection pressures on pro- and
  anti-social punishment. Contrary to the suggestions of previous
  work, we find that anti-social punishment can prevent the evolution
  of pro-social punishment and cooperation under a range of group
  structures. Given that anti-social punishment has now been found in
  all studied extant human cultures, the claims of previous models
  showing the co-evolution of pro-social punishment and cooperation in
  group-structured populations should be re-evaluated. [This is a post-print of an accepted manuscript published in \textit{Journal of Theoretical Biology} 311 (2012) 107--116. The publisher's version is available from http://www.sciencedirect.com/science/article/pii/S002251931200344X.]

\end{abstract}


\end{frontmatter}

\section{Introduction}
\label{secIntro}

Understanding the evolution of individually-costly cooperative
behaviours is a major focus of social evolution theory
\citep{Hamilton:1964:a,Wilson:1975:a,Frank:1998:a,Lehmann:2006:a,West:2007:b,Archetti:2012:a}.
It is now widely appreciated that cooperative behaviours can evolve if
they provide either a direct fitness benefit to the actor during its
lifetime, or if they provide an indirect fitness benefit by helping
other cooperators \citep[e.g.
relatives,][]{Hamilton:1964:a,Lehmann:2006:a,West:2007:b}. %
A great deal of current research is aimed at understanding the
biological mechanisms that provide direct or indirect fitness benefits
in different scenarios \citep{Hammerstein:2003:a,West:2007:b}. In
particular, identifying the mechanisms that provide direct or indirect
fitness benefits to cooperators in large human groups, where genetic
relatedness is typically low, remains an open challenge. This question
has received much attention, due to the fact that humans appear to
cooperate on a much larger scale than other species, and thus
large-scale cooperation seems to be one of the properties that make
human sociality unique.

Cooperation in humans is often framed in terms of the production and
sharing of various public goods. This may take the form of, for
example, sharing food or information with other group members, or
contributing time, energy, and resources to a group
project. Throughout this paper, we focus on social dilemmas that take
the form of \emph{linear public goods games}\footnote{See
\citet{Archetti:2012:a} for a discussion of the differences between
linear and non-linear public goods games and how these affect the
selective dynamics of cooperation.}. In linear public goods games (PGG) there
is an apparent individual advantage to defection, that is, to reaping the
benefits of the public good without paying the individual costs of
contributing to it. We would expect such defectors to be fitter than
cooperators within the same group, and hence for natural selection to
lead to the breakdown of cooperation in a ``Tragedy of the Commons''
\citep{Hardin:1968:a}. However, we see that cooperation is
nevertheless maintained in large human groups despite the apparent
advantage of defection. Explaining this is problematic because many
cases of cooperation in humans occur in large groups of unrelated
individuals.

Punishment behaviours have been widely suggested as a solution to this
quandary. In particular, cooperative individuals may have the option
of punishing defectors. Typically, punishment takes the form of an
actor paying a cost by reducing its own fitness in order to reduce the fitness of
the punishment target, as in the following fitness functions:
\begin{align}
w_d &= 1 + Bx_p -Px_p \\
w_p &= 1 + Bx_p -C -Kx_d,
\end{align}
where $w_d$ is the fitness of defectors, and $w_p$ the fitness of
pro-social punishers (individuals who cooperate and then punish any
group member that defected). In these functions $B$ is a constant
representing the benefit that a single cooperator provides, $C$ the
cost of cooperation, $P$ a constant representing the cost of being
punished, and $K$ the cost of punishing an individual.  %
The constant $1$
represents a baseline fitness in the absence of social
interactions.

The proportions of defectors and pro-social punishers within a group
are denoted by $x_d$ and $x_p$, respectively.  The relative cost of
$P$ and $K$ is subject to some debate, however, empirical evidence from PGG
experiments show similar effects for a range of relative values
\citep{AndersonPutterman06}. In the present paper we assume that the
cost of being punished is partially distributed, and depends upon the
\emph{proportion} of punishers, not the absolute number, in a
group. This is a common assumption in models of the evolution of
punishment \citep{Boyd:1992:a,Boyd:2003:a,Lehmann:2007:c}. We assume
that an individual has a fixed amount of time and energy that can be
spent on acts of punishment, and also \citep[as in prior
work,][]{Boyd:1992:a,Boyd:2003:a} that each punisher punishes every
available target in its group, or at least that their cost of
punishment is proportional to their probability of encountering
punishment targets.  The more individuals there are for an actor
to punish, the less effort the actor can exert on punishing any one
individual, and hence the less absolute damage is inflicted per
punisher per target. %
Note that in most literature, punishment targets are taken to
be defectors, but in the case of anti-social punishment targets may be
cooperators.  %

In the above typical model, %
when pro-social punishers are sufficiently common within a
group, it is individually advantageous for potential defectors to switch
behaviour to cooperating. The condition for this is $Px_p > Kx_d + C$, that is, when
punishers are in sufficient frequency that the cost of punishing and
cooperating (i.e. being a pro-social punisher) is less than the cost
of being punished \citep{Lehmann:2007:c}. Thus, in this simple model
the evolutionarily stable states are either pro-social punishment at
fixation, or defection at fixation. Note that this result holds even
in a well-mixed population, and so would be applicable to large groups
of unrelated individuals \citep{Boyd:1992:a}.

There are, however, at least three problems with this as a mechanism
for the maintenance of cooperation in human groups. The first is that
because both pro-social punishment and defection are equilibria, why
would we expect natural selection to lead to one rather than the other
\citep{Boyd:2003:a}? It has been argued that group selection \citep[in
the broad sense,][]{Okasha:2006:a} should promote the pro-social
punishment equilibrium \citep{Boyd:2003:a,Bowles:2004:a}, because this
increases the mean fitness of group members. The second problem is
that pro-social punishment may not actually be stable under mutation
if cooperation and punishment are not perfectly linked traits
\citep{Lehmann:2007:c}. This is because pro-social punishers may be
slowly replaced by non-punishing cooperators that do not pay the cost
of punishing, and who are not themselves punished \citep[the second-order free-riding problem,][]{Colman:2006:a}. As pro-social
punishers decline in frequency due to the accumulation of non-punish
mutations, defection may again become advantageous. The third problem
is that there is no reason to suppose that only defectors can be
targets of punishment: defectors may have the option of punishing
cooperators, as evidenced by recent empirical work
\citep{Herrmann:2008:a}.  Exploring the consequences of this third
problem, termed {\em anti-social punishment} (ASP) motivates the
present investigation.

Recent theoretical work \citep{Rand:2010:a,Rand:2011:a} has
suggested that anti-social punishment can thwart the evolution of
pro-social punishment and cooperation. However, no study has formally
addressed anti-social punishment in group-structured populations, even
though such population structures are frequently modelled when
considering the evolution of pro-social punishment
\citep{Boyd:2003:a,Boyd:2010:a,Bowles:2004:a,Lehmann:2007:c}. 
Group-structured populations are an essential component of
cultural group selection, and of recent arguments about punishment
promoting group-beneficial cultural norms in humans
\citep{Boyd:2003:a,Gintis:2003:a,Mathew:2011:a,Sober:1998:a}. Thus
such populations warrant an explicit model in order to determine
whether the inclusion of anti-social punishment affects the claims of
these works. For example, some previous work has verbally argued that
the presence of group-structure should be expected to favour
pro-social punishment even when anti-social punishment is present
\citep{Rand:2010:a,Rand:2011:a}. According to this view,  we
might not expect anti-social punishment to have any effect in
group-structured populations. %
Clearly, the implications for the understanding of
the proximate mechanisms promoting cooperation in
humans motivate a formal examination of this case.

In this paper, we formally address the evolution of anti-social
punishment in group-structured populations, focusing on the conditions
under which it prevents the evolution of pro-social punishment. We
also consider the effects of non-punishing cooperator and defector
mutations.

\section{A model of the evolution of pro- and anti-social punishment in group-structured populations}
\label{secModel}
We consider here a population of individuals that live and reproduce in social groups for a number of generations. In each generation, the following two-phase social interaction occurs and determines fitness within the group. Firstly, each individual either cooperates by paying an individual cost to contribute to a public good, or defects by not contributing. All group members then receive an equal share of the benefit of this good, regardless of whether they cooperated in its production or not. Then, in the second stage individuals have the option of punishing other group members, based on how they behaved in the first stage. We model here the evolution of four behavioural strategies: cooperate but do not punish (non-punishing cooperate); defect but do not punish (non-punishing defect); cooperate and punish all individuals that defected (pro-social punisher); defect and punish all individuals that cooperated (anti-social punisher). The fitness of these four types within a single group is given, respectively, by the following fitness functions:
\begin{align}
w_c &= 1 + B(x_c+x_p) -C -Px_a \label{eqnWc}\\
w_d &= 1 + B(x_c+x_p) -Px_p \label{eqnWd}\\
w_p &= 1 + B(x_c+x_p) -C -K(x_d+x_a) -Px_a \label{eqnWp}\\
w_a &= 1 + B(x_c+x_p) -K(x_c+x_p) -Px_p \label{eqnWa}.  
\end{align}
 
Here, $x_c$, $x_d$, $x_p$ and $x_a$ are the proportions of non-punishing cooperators, non-punishing defectors, pro-social punishers and anti-social punishers, respectively. As in the previous section, $B$ is a constant representing the benefit that a single cooperator provides, and $C$ is a constant representing the cost of cooperation. $K$ and $P$ are constants representing the cost of punishing and being punished, respectively. The constant $1$ represents a baseline fitness in the absence of social interactions. These fitness functions are based on those commonly used to model the evolution of pro-social punishment (e.g. \citealp{Boyd:2003:a,Lehmann:2007:c}), with the addition here of anti-social punishment. Note that in this model, pro-social punishers punish both types of defectors, i.e. non-punishing defectors and anti-social punishers. Likewise, anti-social punishers punish both types of cooperators, i.e. non-punishing cooperators and pro-social punishers.  

The number of individuals of type $i$, $n_i$, within a group changes deterministically each generation according to the following difference equation:
\begin{align}
n_i(t+1) = n_i(t) + n_i(t)w_i, \label{eqnReproduction}
\end{align}
where $t$ refers to the current generation, and $w_i$ is the fitness of type $i$ within the group at the current generation, as given by Equations~\ref{eqnWc}--\ref{eqnWa}. Fractional parts are maintained throughout. We assume that reproduction is asexual, and that genotypes are haploid with a single locus determining behaviour. Note that in Equation~\ref{eqnReproduction} all individuals of the same type (strategy) within any one group have the same fitness, and reproduce by the same amount. Thus, within groups we do not need to explicitly track each individual, but can simply track type densities. 

The above fitness functions describe social interactions within single groups. At the metapopulation level, we model groups formed by random sampling of $n$ individuals from a global migrant pool of size $N$. This sampling is done without replacement, according to a multivariate hypergeometric distribution. Reproduction and selection then occurs deterministically within these groups for $T$ generations, according to Equation~\ref{eqnReproduction}. Note that there is no local density regulation in Equation~\ref{eqnReproduction}, thus, different groups may grow to different sizes over the course of the $T$ generations. After $T$ generations dispersal occurs (ecologically, dispersal could be triggered by depletion of a resource patch, for example). During the dispersal stage, a new migrant pool is formed by summing the absolute type densities across all groups. Groups that have grown to a larger size will make up a larger fraction of the new pool, representing a form of global competition between groups. Global population regulation then occurs by proportionality rescaling the migrant pool back to size $N$. The number of individuals of a type after population regulation is computed by calculating the proportion of that type in the migrant pool and multiplying by $N$, rounding the result to remove fractional parts and produce an integer number of individuals.

Each individual in the migrant pool undergoes mutation with probability $\mu$. If chosen for mutation, an individual's genotype is changed randomly to one of the other three types with equal probability. The individuals in the migrant pool then form the next generation of groups (i.e. colonise a new set of resource patches), as previously described. This process of group formation and dispersal then continues for a number of cycles, $G$. We simulate the model by the following procedure:

\begin{enumerate}
	\item \textbf{Initialisation}: form a migrant pool of $N$ individuals with $N_c$ non-punishing cooperators, $N_d$ non-punishing defectors, $N_p$ pro-social punishers and $N_a$ anti-social punishers.
	\item \textbf{Group formation}: form $\left\lfloor N/n \right\rfloor$ groups of size $n$ by random sampling from the migrant pool without replacement (where $\left\lfloor \right\rfloor$ denotes the mathematical floor function).  
	\item \textbf{Reproduction and selection within groups}: iterate equation~\ref{eqnReproduction} $T$ times for each group (see text). 
	\item \textbf{Dispersal}: all individuals leave their groups to form a new migrant pool. The migrant pool is then rescaled back to size $N$, keeping the proportion of types the same. Fractional parts are rounded. 
	\item \textbf{Mutation}: each individual in the migrant pool undergoes mutation with probability $\mu$ (see text). 
	\item \textbf{Iteration}: repeat from step 2 for $G$ cycles. 
\end{enumerate}

The population structure described above is based on the Haystack
model
\citep{Smith:1964:a,Cohen:1976:a,Wilson:1987:a,Sober:1998:a,Bergstrom:2002:a,Fletcher:2004:a,Fletcher:2007:a,Powers:2011:a},
where we allow both the size of social groups when they are founded,
and the frequency of dispersal, to be parameterised. When $T=1$ and
dispersal occurs every generation, this corresponds to the
well-studied trait-group model of \citeauthor{Wilson:1975:a}, \citetext{\citeyear{Wilson:1975:a}; c.f. \citealt{Hamilton:1975:a,Michod:1983:a,Nunney:1985:a,Smith:1995:a,Pepper:2000:a,Okasha:2006:a,Santos:2008:a}}. This provides us with a simple model that allows the ecological and demographic factors affecting between-group variance, and hence the strength of group selection, to be varied \citep{Sober:1998:a}.

\section{Results}
\label{secResults}
In this section we show how two demographic factors, group size and
dispersal frequency, affect whether selection will favour pro- or
anti-social punishment, or no punishment at all. We first present
analytical results from a simpler version of the model in which pairs
of strategies compete, and dispersal frequency is fixed with groups
dispersing after every generation. This population structure
corresponds to Wilson's \citeyearpar{Wilson:1975:a} trait-group
model. It also parallels the anonymous single-shot public goods games
used in behavioural economics experiments, particularly those used to
study strong reciprocity \citep{Fehr:2002:a}. We then go on to present
results from the full simulation model in which all four strategies
are present simultaneously, and both dispersal frequency as well as
group size is varied. Less frequent dispersal ($T>1$) corresponds to a
Haystack population structure, and is an evolutionary analogue of
repeated economic public goods games. %

\subsection{Analytical results}
We explore first the evolutionary dynamics where only two strategies
are present in the population at any one time, and where dispersal
occurs every generation $(T=1)$. We first note that the fitness of
each type is frequency-dependent
(Equations~\ref{eqnWc}--\ref{eqnWa}). In particular, the fitness of
both pro- and anti-social punishers is positive
frequency-dependent. This is because the total cost of punishing
decreases as punishers become more common (since less acts of
punishment are required), while the total cost of being punished
increases (due to more individuals performing punishing acts). Thus,
although neither type of punishment may easily be able to invade from
rarity (though see Discussion, Section~\ref{disc.sec}), %
they can be
selected for from higher initial frequencies
\citep{Lehmann:2007:c}. We therefore focus on how the initial
proportion of a strategy required for it to be selectively favoured
changes with respect to group size. In addition to the analytical
results, we verified each threshold frequency derived below
numerically in the simulation model.

\subsubsection{Non-punishing cooperators vs. non-punishing defectors}
\label{coopdef.sec}
We first consider pairwise competition between non-punishing cooperators and non-punishing defectors. The fitness of each type, given a group containing $j$ cooperating co-players, is:%
\begin{eqnarray}
w_c(j)&=&B\frac{j+1}{n}-C+1\nonumber\\
&=&B\frac{j}{n}+\frac{B}{n}-C+1\nonumber\\
w_d(j)&=&B\frac{j}{n}+1
\end{eqnarray}
To calculate whether cooperation increases or decreases in proportion, we need to know the probability distribution $\mathbb{P}_n(j)$ for an individual to be placed in a group with $j$ cooperators, given that $x_c$ is the global frequency of cooperators. In our model, we assume that groups are formed by a random sampling process. In the full model, this sampling occurs without replacement from a finite population according to a hypergeometric distribution. However, for ease of analysis we derive the analytical results in this section by sampling groups from an infinite population with replacement; this does not significantly alter the qualitative behaviour of the results. We thus model group formation in this section by sampling from a binomial distribution.

\begin{figure}[htb]
\centering
\subfloat[]{\label{figCvsD}\includegraphics[scale=0.2]{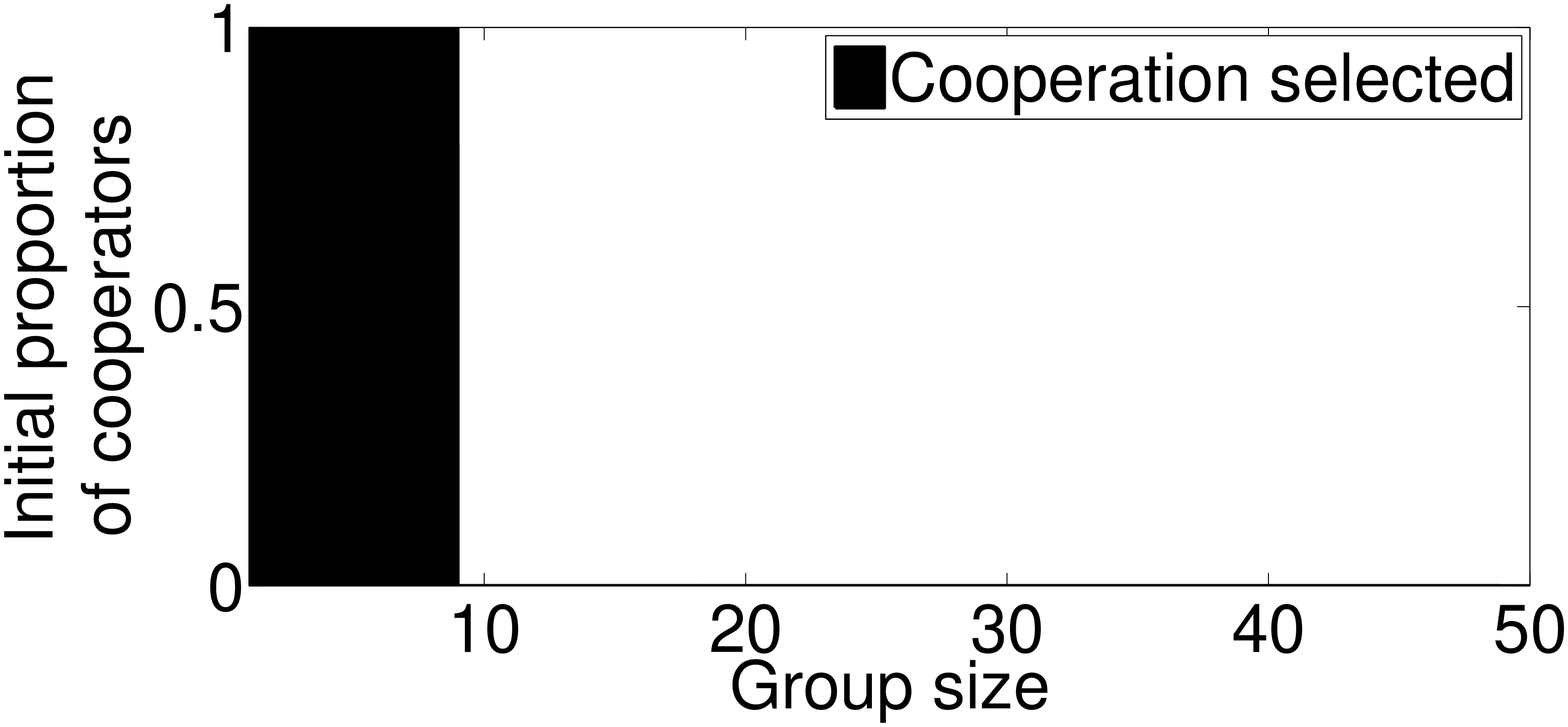}}
 \subfloat[]{\label{figPSPvsD}\includegraphics[scale=0.2]{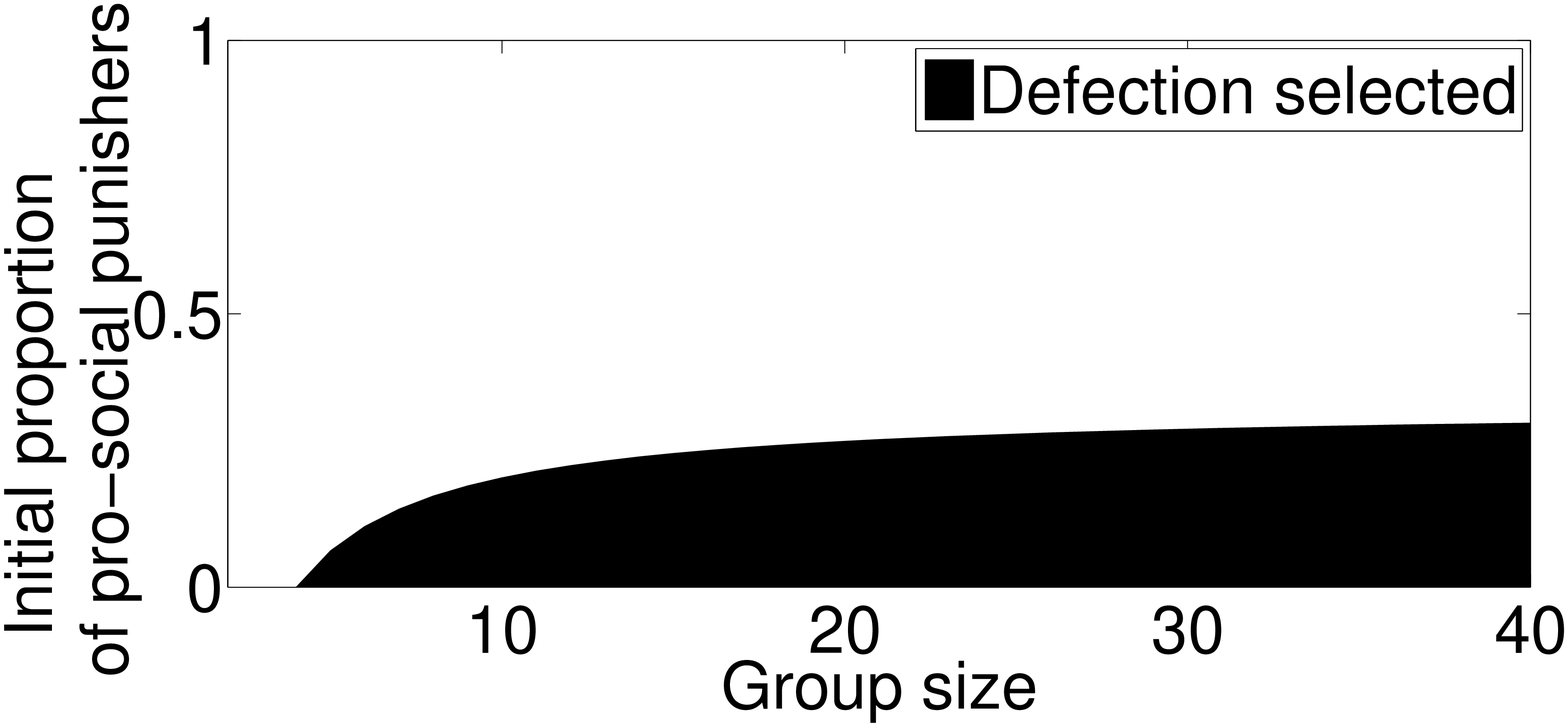}}
\\ \subfloat[]{\label{figASPvsC}\includegraphics[scale=0.2]{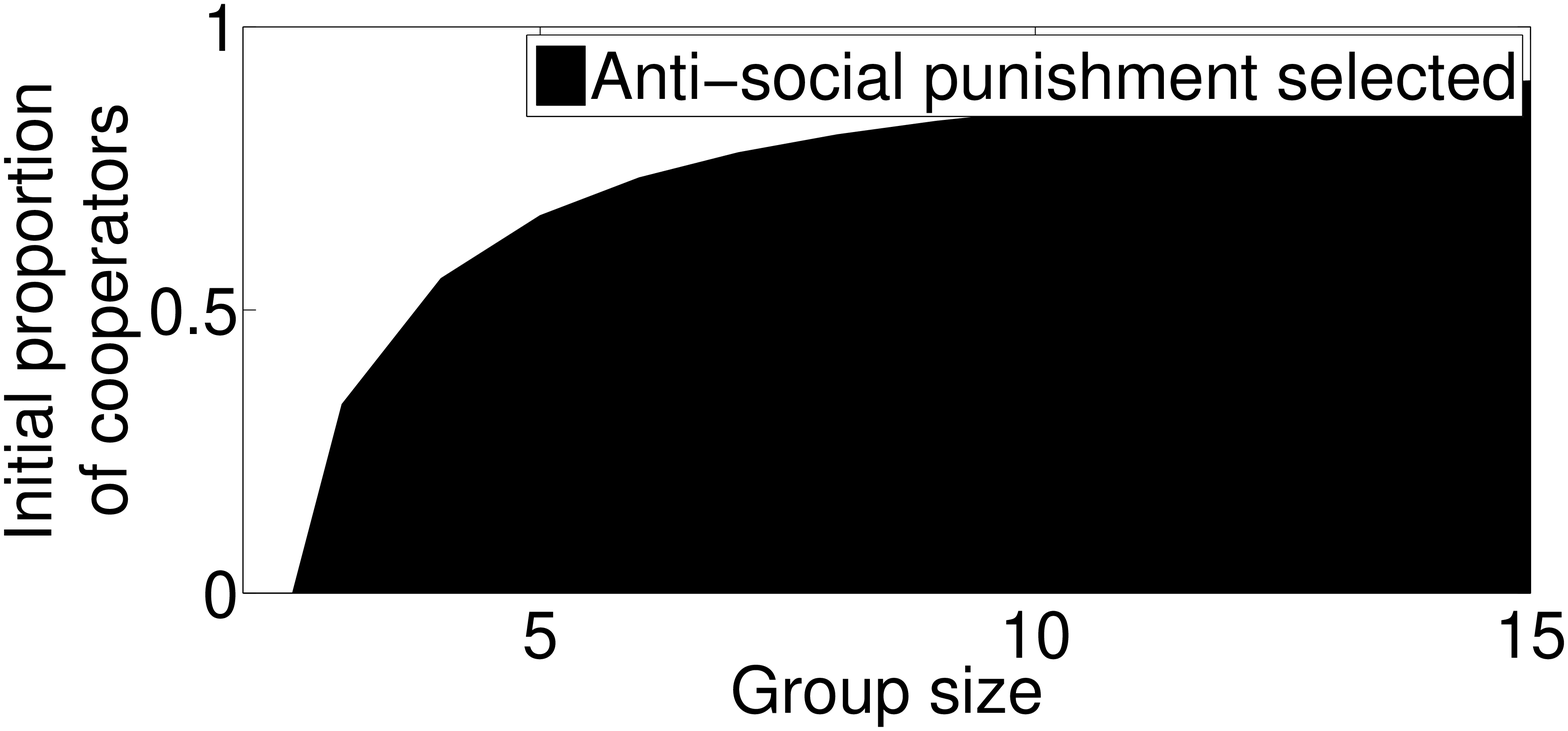}}
 \subfloat[]{\label{figASPvsPSP}\includegraphics[scale=0.2]{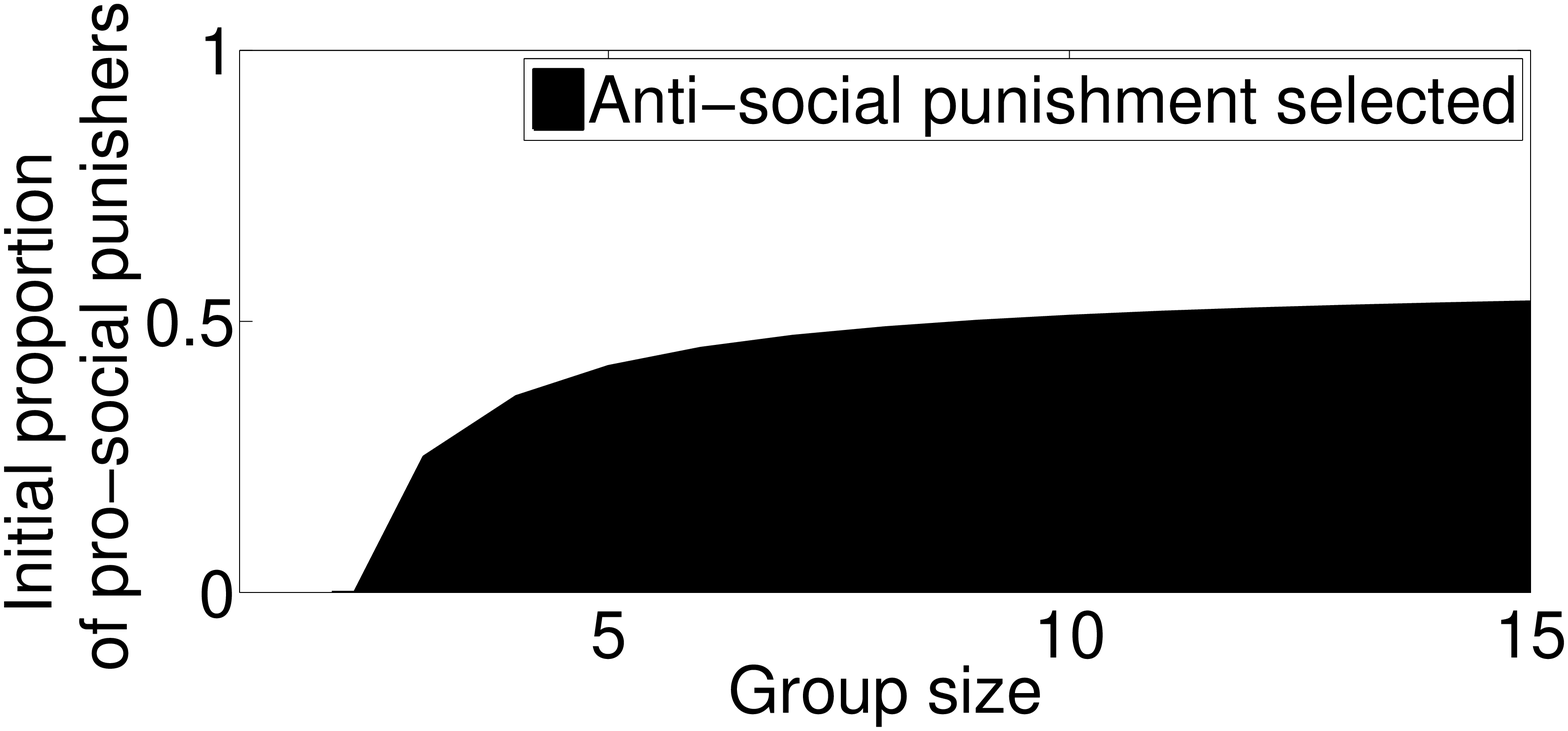}}
\caption{Analytical results of pairwise competition between
  strategies. Groups are formed by binomial sampling from the global
  population and disperse every generation (trait-group
  model). For this plot, parameters other than group size are fixed to the values used later in the simulations (B=0.9, C=0.1, K=0.1,
  P=0.5), c.f. main text. a) Non-punishing cooperators vs. non-punishing defectors. b) Pro-social
  punishers vs. non-punishing defectors. c) Anti-social punishers
  vs. non-punishing cooperators. d) Anti-social vs. pro-social punishers.}%
\label{figPairwise}
\end{figure}

Let $g_c$ and $g_d$ be the fitness of the cooperators and defectors \emph{after} the groups have been formed. We can calculate them by
\begin{eqnarray}
g_c&=&\sum^{n-1}_{j=0}\mathbb{P}_n(j)w_c(j)\nonumber\\
     &=& \sum^{n-1}_{j=0}\binom{n-1}{j} x_c^j (1-x_c)^{n-1-j}w_c(j)\nonumber\\
     &=& x_cB\frac{n-1}{n}+\frac{B}{n}-C+1\nonumber\\
g_d&=&\sum^{n-1}_{j=0}\mathbb{P}_n(j)w_d(j)\nonumber\\%changed a to d
    &=&\sum^{n-1}_{j=0}\binom{n-1}{j} x_c^j (1-x_c)^{n-1-j}w_d(j)\nonumber\\
&=&x_cB\frac{n-1}{n}+1,
\end{eqnarray}
where $x_c$ is the \emph{global} proportion of cooperators.
We see that cooperation increases in frequency in the population when  $g_c>g_d$,%
 which occurs when
\begin{equation}
\frac{B}{n}>C.
\end{equation}
This is a standard result for the evolution of weak altruism in linear public goods games, i.e.
social traits that increase the \emph{absolute} fitness of the actor
but increase the absolute fitness of other group members by even more
\citep{Wilson:1975:a,Wilson:1979:a,Wilson:1990:a,Nunney:1985:a,Szathmary:2011:a}. Because
all group members receive the benefit of cooperation, including the
actor, then cooperation is selected for whenever the actor's share of
the benefit exceeds the cost to itself. This
depends on group size and is given by $B/n>C$ (Figure~\ref{figCvsD}). This type of cooperation corresponds to what \citet{Pepper:2000:a} terms a whole-group trait. Biological examples of such whole-group traits include siderophore production in bacteria \citep{Griffin:2004:a}, and the efficient use of shared resources through lower consumption rates \citep{Pfeiffer:2001:a,Kreft:2004:a,Killingback:2006:a}. Whole-group traits also parallel the setup of
economic public goods games \citep{Fehr:2002:a,Herrmann:2008:a}. For such traits a smaller group size favours cooperation because a cooperator
receives a larger share of the benefits of its actions. In this model
whenever $B/n>C$, cooperation fixes in the global population
irrespective of starting frequency, assuming deterministic evolution. For $B/n<C$ defection fixes. An
equivalent way of thinking about this is that for whole-group traits
(such as public goods), the relatedness of actor to recipients in this
population structure is $1/n$, assuming an infinite global population
size \citep{Pepper:2000:a}. Relatedness here is taken to mean a genetic correlation between actor and recipients at the locus for cooperation \citep{Foster:2006:a,West:2007:b}. The condition $B/n>C$ is then an instance of Hamilton's rule \citep{Hamilton:1964:a}.

\subsubsection{Pro-social punishers vs. non-punishing defectors}
We next investigate the evolutionary dynamics of pro-social punishers compared to defectors. Recall that pro-social punishers are cooperators that identify individuals that defected, and then pay a fitness cost $K$ to reduce the fitness of defectors in their group by $P$. From this we can derive the fitness of both strategies, given that they are placed in a group containing $j$ pro-social punishers
\begin{eqnarray}
w_p(j)&=&B\frac{j+1}{n}-C-K\frac{n-(j+1)}{n}+1\nonumber\\
w_d(j)&=&(B-P)\frac{j}{n}+1 %
\end{eqnarray}
As before, we assume that the distribution of strategies to groups is binomial; $\mathbb{P}_n(j)$ is then the probability of being placed in a group with $j$ pro-social punishers, given that groups are of size $n$. From this, we calculate the fitnesses of each individual: 
\begin{eqnarray}
g_p&=&\sum^{n-1}_{j=0}\mathbb{P}_n(j)w_p(j)\nonumber\\
     &=&\sum^{n-1}_{j=0}\binom{n-1}{j} x_p^j (1-x_p)^{n-1-j}w_p(j)\nonumber\\
     &=& x_p(B+K)\frac{n-1}{n}+\frac{B}{n}-C-K\frac{n-1}{n}+1\nonumber\\
g_d&=&\sum^{n-1}_{j=0}\mathbb{P}_n(j)w_d(j)\nonumber\\
     &=&\sum^{n-1}_{j=0}\binom{n-1}{j} x_p^j (1-x_p)^{n-1-j}w_d(j)\nonumber\\
&=&x_p(B-P)\frac{n-1}{n}+1,
\end{eqnarray}
where $x_p$ is the global frequency of pro-social punishers.
Pro-social punishment increases in proportion when $g_p>g_d$, or when 
\begin{equation}
x_p>x_p^*=\frac{Cn-B}{(n-1)(P+K)}+\frac{K}{P+K},
\end{equation}
where $x_p^*$ is the proportion of punishers needed in the population
for punishment to be selectively favoured. $x_p^*$ is horizontally asymptotic,
with respect to group size $n$, with the asymptote at
$x_p^*=\frac{C+K}{P+K}$%
. Figure~\ref{figPSPvsD} plots this critical value as function of $n$,
the group size, using the parameters ($B=0.9$, $C=0.1$, $K=0.1$,
$P=0.5$) that we later use in the simulations; we discuss robustness of the numerical results with respect to these parameters in Section~\ref{sensitivitySec}.  This result agrees with Equation 11 of
\citet{Lehmann:2007:c} (noting that we include the effect of a
punisher on itself) for the case of randomly formed groups that
disperse every generation, and applies where pro-social punishment and
cooperation are perfectly linked traits. We relax these assumptions in
the simulation model (Section~\ref{secSimRes}).

\subsubsection{Non-punishing cooperators vs. anti-social punishers}
We now turn to investigate the dynamics of anti-social punishers in competition with non-punishing cooperators. In particular, we focus on whether anti-social punishment can prevent the invasion of cooperation even when $B/n>C$, i.e. even when the actor gains in absolute fitness terms from cooperating. The fitnesses in a group containing $j$ cooperators are:
\begin{eqnarray}
w_c(j)&=&B\frac{j+1}{n}-C-P\frac{n-(j+1)}{n}+1\nonumber\\
w_a(j)&=&(B-K)\frac{j}{n}+1
\end{eqnarray}
Let $g_c$ and $g_a$ be the fitness of non-punishing cooperators and anti-social punishers respectively. Again let $\mathbb{P}_n(j)$, the probability of an individual being placed in a group with $j$ cooperators, have a probability mass function of a binomially distributed random variable. %
 Then
\begin{eqnarray}
g_c&=&\sum^{n-1}_{j=0}\mathbb{P}_n(j)w_c(j)\nonumber\\
&=&\sum^{n-1}_{j=0}\binom{n-1}{j} x_c^j (1-x_c)^{n-1-j}w_c(j)\nonumber\\
     &=& x_c(B+P)\frac{n-1}{n}+\frac{B}{n}-C-P\frac{n-1}{n}+1\nonumber\\
g_a&=&\sum^{n-1}_{j=0}\mathbb{P}_n(j)w_a(j)\nonumber\\
&=&\sum^{n-1}_{j=0}\binom{n-1}{j} x_c^j (1-x_c)^{n-1-j}w_a(j)\nonumber\\
&=&x_c(B-K)\frac{n-1}{n}+1,
\end{eqnarray}
where $x_c$ is the global proportion of cooperators.
Cooperation thus increases in proportion when: 
\begin{equation}
x_c>x_c^*=\frac{Cn-B}{(n-1)(P+K)}+\frac{P}{P+K}.
\end{equation}
Critically, we see that cooperation can also \emph{decrease} in proportion, even when $B/n-C>0$. Figure~\ref{figASPvsC} plots the critical value $x_c^*$ against group size $n$. This has a horizontal asymptote at $x_c^*=\frac{C+P}{P+K}$.

\subsubsection{Anti-social vs. pro-social punishers}
Finally, we consider the dynamics of a population containing both pro- and anti-social punishers. As before, we let $w_p(j)$, and $w_a(j)$ be the fitness of a pro-social punisher and an anti-social punisher in a group of size $n$, containing $j$ pro-social punishers (excluding self). Then
\begin{eqnarray}
w_p(j)&=&B\frac{j+1}{n}-C-(P+K)\frac{n-(j+1)}{n}+1\nonumber\\
w_a(j)&=&(B-K-P)\frac{j}{n}+1
\end{eqnarray}
Let $g_p$ and $g_a$ be the fitness of each type after they have been placed into groups. The groups are again formed binomially. Then,
\begin{eqnarray}
g_p&=&\sum^{n-1}_{j=0}\mathbb{P}_n(j)w_p(j)\nonumber\\
&=&\sum^{n-1}_{j=0}\binom{n-1}{j} x_p^j (1-x_p)^{n-1-j}w_p(j)\nonumber\\
&=& x_p(B+P+K)\frac{n-1}{n}+\frac{B}{n}-C-(P+K)\frac{n-1}{n}+1\nonumber\\
g_a&=&\sum^{n-1}_{j=0}\mathbb{P}_n(j)w_a(j)\nonumber\\
&=&\sum^{n-1}_{j=0}\binom{n-1}{j} x_p^j (1-x_p)^{n-1-j}w_a(j)\nonumber\\
&=&x_p(B-K-P)\frac{n-1}{n}+,
\end{eqnarray}
where $x_p$ is the global proportion of pro-social punishers. 
Thus, punishment increases in proportion in the global population when 
\begin{equation}
x_p>x_p^*=\frac{Cn-B}{2(n-1)(P+K)}+\frac{1}{2}.
\end{equation}
This has a horizontal asymptote at $x_p^*=\frac{C+K+P}{2(P+K)}$. The critical initial frequency for pro-social punishment to be selected is plotted with respect to group size in Figure~\ref{figASPvsPSP}.

\subsubsection{Pro-social punishment vs. cooperation, anti-social punishment vs. defection, and the second-order free-riding problem}
In the absence of other strategies, pro-social punishment is neutral with non-punishing cooperate. This is because in such a case there are no individuals to be punished, and hence the total cost of punishing is zero. Likewise, anti-social punishment is neutral with non-punishing defection in the absence of other strategies. However, all four strategies may always be present in above-zero frequency in a population due to mutation. In such cases the effects of pro-social punishment, in terms of reducing the proportion of defectors and anti-social punishers in the group, are shared equally by both pro-social punishers and non-punishing cooperators. Non-punishing cooperators, however, would not pay the cost of punishing and so would be expected to be fitter \citep[the second-order public goods problem,][]{Colman:2006:a,Eldakar08}. Similarly the effects of anti-social punishment, in terms of reducing the frequency of competing individuals with the cooperative trait, are felt equally by both anti-social punishers and non-punishing defectors in the same group.  Non-punishing defectors, however, do not pay the cost of punishment. Thus, since in this model neither type of punishment differentially benefits punishers within a group, we might expect both types of punisher to be replaced by their non-punishing counterparts. Hence, we would not expect either type of punishment to be evolutionarily stable when non-punishing mutants are introduced. We investigate this through simulation in the next section.

\subsection{Simulation results}
\label{secSimRes}
For the primary simulation results below, we fix the migrant pool size
at $N=500$, the benefit of cooperation at $B=0.9$, the cost of
cooperation at $C=0.1$, the cost of being punished at $P=0.5$, the
cost to the punisher at $K=0.1$, and the mutation rate at $\mu=0.01$. We
vary the group size, $n$, and the number of generations before
dispersal, $T$. We thus focus on the demographic factors that affect
selection on punishment. We return to examining parameter
sensitivity after presenting the primary results
(Section~\ref{sensitivitySec}). %

As the analytic results above show, outcomes are sensitive to initial
conditions, since selection pressures depend upon the frequencies of
all four strategies (Equations~\ref{eqnWc}--\ref{eqnWa}). As a
consequence, although a type may not be likely to invade when rare, it
may be maintained or increase in frequency by selection when established
at sufficient frequency. %
In this study, we focus on the initial
condition where all four strategies are present in equal frequency. We
address the validity of this assumption in the Discussion
(Section~\ref{disc.sec}). As we show below, from this state pro-social
punishment and cooperation will be selected \emph{against} in a
well-mixed population. We therefore investigate how the presence of various
types of group structure changes selection pressures on the
maintenance of both types of punishment.

\subsubsection{Well-mixed population}

\begin{figure}[!htb]
\centering
\subfloat[]{\label{figFMEqualProp}\includegraphics[scale=0.32]{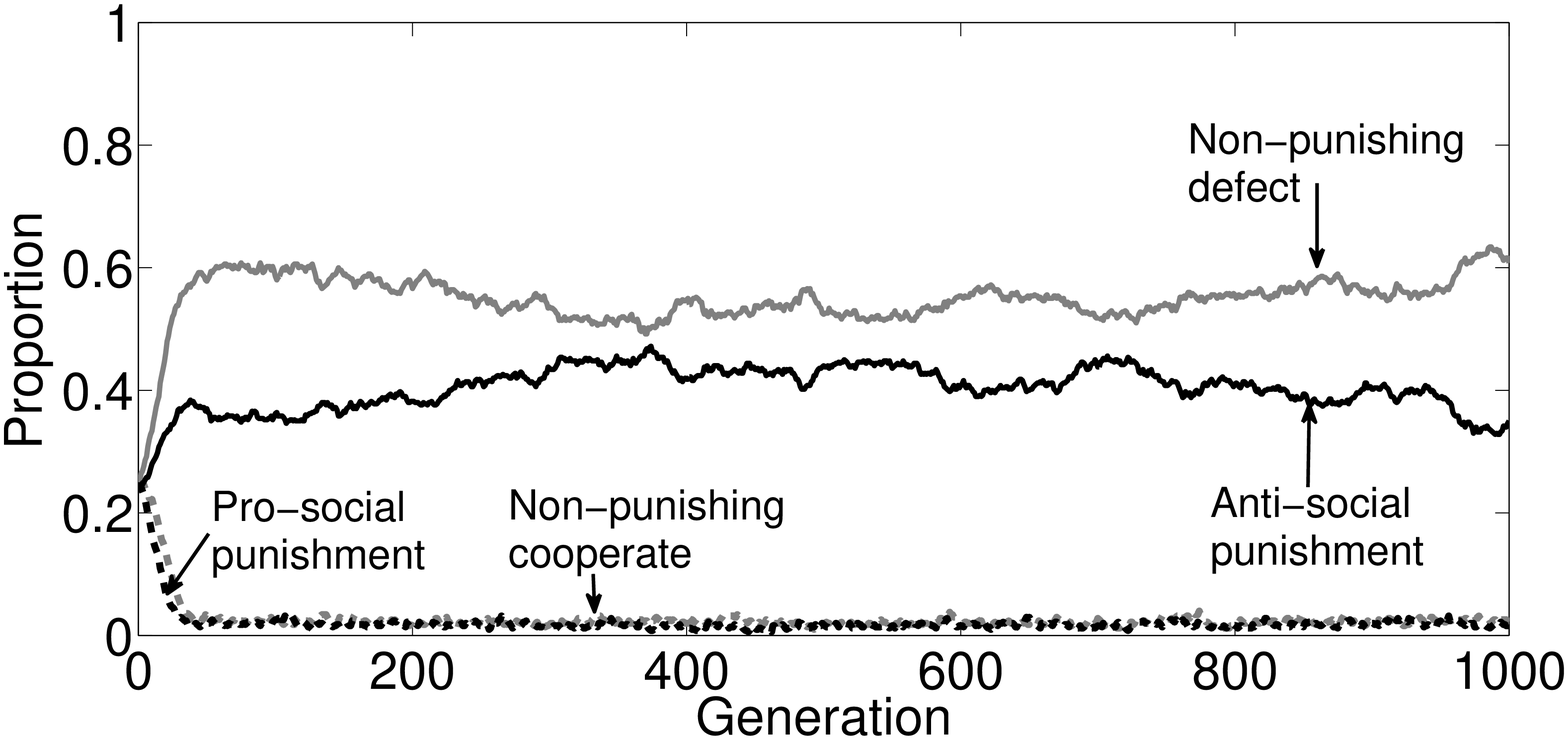}}\\ \subfloat[]{\label{figFMPSPFixed}\includegraphics[scale=0.32]{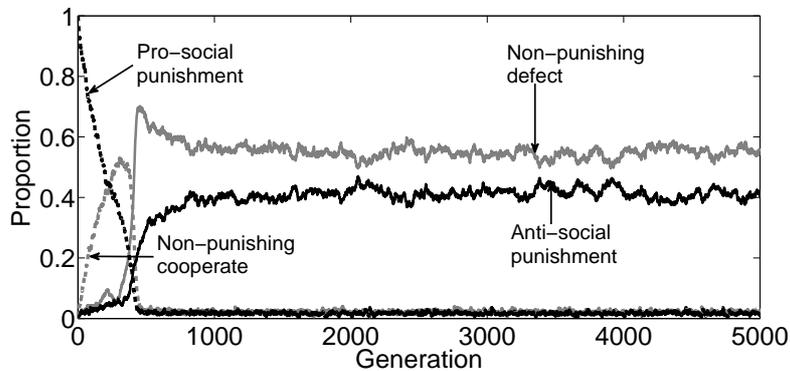}}
\caption{Simulation results of competition between pro-social
  punishers, anti-social punishers, non-punishing cooperators, and
  non-punishing defectors in a \emph{well-mixed population} with mutation. a)
  All types initially present in equal frequency, showing the
  selective advantage of defection from this initial frequency. b)
  Pro-social punishment initially at fixation. Although stable as a
  strategy against invasion by defectors only (not shown), the
  increase in non-punishing cooperator mutants creates a second-order
  Tragedy of the Commons in which first pro-social punishment, and
  subsequently cooperation, collapses.
} %
\label{figFM}
\end{figure}

We first investigate the effects of mutation in a well-mixed
population. In the present model, we do this by setting $n=N=500$ and
$T=1$. With all four strategies started in equal frequency, both
cooperate and pro-social punishment are driven extinct, apart from
reintroduction by recurring mutation (fig.~\ref{figFMEqualProp}). If
anti-social punishment was neutral with non-punishing defect, we would
expect both types to reach a proportion of approximately 50\% under
mutation. However, we see that anti-social punishment is in fact
weakly selected against, being held at a frequency of around
40\%. This is because anti-social punishers are paying the cost of
punishing the few cooperators and pro-social punishers that are being
reintroduced by mutation. The total cost of these acts of punishment
is not large, however, because there are few individuals to punish and
cooperators are selected against in a well-mixed population even
without anti-social punishment (see analytical results,
Section~\ref{coopdef.sec}).

We next consider the case where pro-social punishment is initially fixed in the population. If only defectors (and anti-social punishers) arose by mutation, then pro-social punishment would be stable. However, we also allow for the possibility of non-punishing cooperators. In this situation, non-punishing cooperator mutants increase in frequency towards 50\% (fig.~\ref{figFMPSPFixed}), as would be expected if pro-social punishment was neutral with non-punishing cooperation. However, the resulting decline in the frequency of pro-social punishment under mutation creates a second-order Tragedy of the Commons. That is, as pro-social punishment drops in frequency acts of punishment become too rare to prevent defection from being favoured; the condition for punishment to prevent defection in a well-mixed population ($Px_p > Kx_d + C$) is then no longer met. The increase in non-punishing cooperators thus creates a selective environment that favours defection. This illustrates that pro-social punishment need not be stable in a well-mixed population when non-punishing mutants are introduced. We also note that this result holds even in the absence of anti-social punishment. Anti-social punishment thus plays no selective role in a well-mixed population, and is maintained at a frequency below that expected under neutrality with non-punishing defect.

\subsubsection{Effect of group size}
We now consider the effects of group structure on the selective
dynamics. We first vary group size, holding $T=1$. With all four
strategies started in equal frequency, we find that cooperation and
pro-social punishment fix in the population for group sizes $n<B/C$.
With $B=0.1$ and $C=0.9$, in our model this is a group size less than
9. This is the classic result for randomly formed, single generational
groups, \emph{without punishment} (see analytical results). Thus, in
this case allowing both cooperators and defectors to punish gives the
same outcome as if neither type could punish. However, if the option
of pro-social punishment is removed (by replacing all pro-social
punishers with non-punishing cooperators in the initial condition),
but anti-social punishment is maintained, we find that the condition
for cooperation to evolve is more stringent than
$n<B/C$. Specifically, we found that cooperation only evolved for a
group size less than 7. On the other hand, if anti-social punishment
is removed from the initial condition (by replacing all anti-social
punishers with non-punishing defectors), then cooperation and
pro-social punishment can evolve even for group sizes where
$n>B/C$. In terms of the parameters used in this study, pro-social
punishment and cooperation reliably evolved in group sizes below 14,
i.e. above the threshold of $B/C=n=9$.

Importantly, when $n>B/C$ then a whole-group trait that provides a
fixed amount of benefit switches from being weakly to strongly
altruistic \citep{Pepper:2000:a}, or becomes ``altruistic'' in the
sense used by \citet{Hamilton:1964:a} and more recently advocated by
\citet{Lehmann:2006:a} and \citet{West:2007:b}.  Such cooperation, when considered outside the context of punishment, causes an absolute reduction in the lifetime fitness of the actor. This illustrates that pro-social punishment can allow a cooperative trait that would otherwise be strongly altruistic to evolve in randomly formed groups. This occurs because pro-social punishment modifies the direct costs and benefits of cooperation, so that it is effectively no longer altruistic \citep{Boyd:1992:a,Lehmann:2007:c}.

We also find that pro-social punishment and non-punishing cooperation
form a stable polymorphic equilibrium in such cases, even in the
presence of recurrent mutations. That is, the second-order Tragedy of
the Commons does not occur in such group-structured populations. The
reason for this, and the maintenance of pro-social punishment and
cooperation even when $n>B/C$, is due to the fact that the population
structure provides localised interactions. Specifically, mean fitness
is lower in groups containing a greater proportion of defectors
(Equations~\ref{eqnWc}--\ref{eqnWa}). Thus after dispersal such groups will make up a smaller fraction of the migrant pool, since the groups will contain fewer individuals relative to
other groups after one iteration of
Equation~\ref{eqnReproduction}. For this to occur, different groups
must contain different proportions of defectors when they are
formed. That is, there must be variance in the composition of groups
\citep{Wilson:1975:a}. Since the groups are formed by random sampling,
a smaller group size provides greater variance. Thus, smaller group
sizes provide stronger selection against defection and anti-social
punishment. %

The results of this section illustrate that either type of punishment can prevent the evolution of a behaviour which would otherwise be selected given the population structure. They also illustrate that when both types of punishment are started in equal frequency their effects can be canceled out, with the population structure again becoming the determinant of selection on cooperation.

\subsubsection{Effect of dispersal frequency}

\begin{figure}[!htb]
\centering
\subfloat[]{\label{figTvsGSnoASP}\includegraphics[scale=0.32]{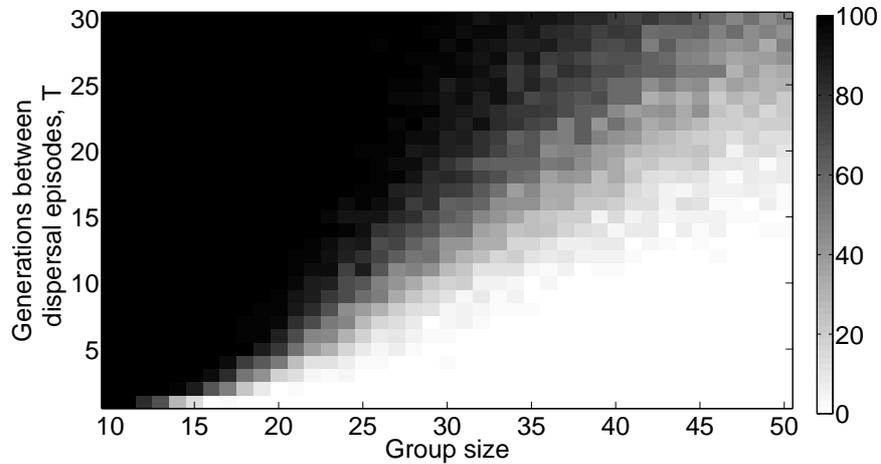}}\\
\subfloat[]{\label{figTvsGSwithASP}\includegraphics[scale=0.32]{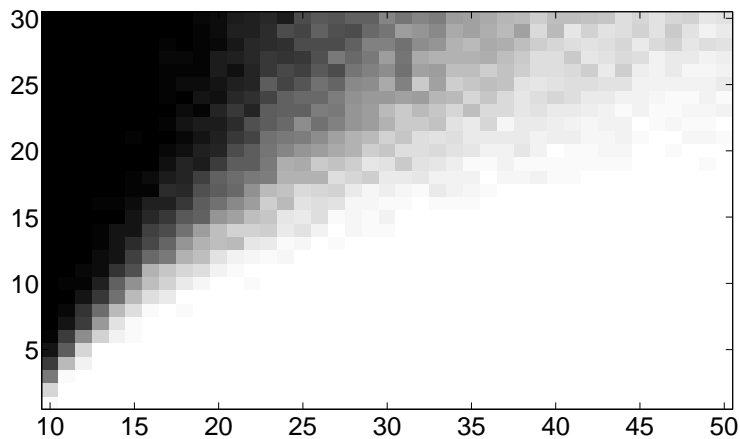}}
\caption{Effect of varying both group size and dispersal
  frequency. The shading indicates the number of simulation runs, out
  of 100, in which selection favoured pro-social punishment and
  cooperation after 1000 cycles. a) Without anti-social punishment; b)
  with anti-social punishment. Smaller group size increases variation
  in group composition, thus providing stronger selection against
  defection. Less frequent dispersal favours pro-social punishment by
  creating an equilibrium selection process. However, the addition of
  anti-social punishment reduces the largest group size under which
  pro-social punishment and cooperation are
  selected. Initial conditions for this simulation are given in the text.} %
\end{figure}

We next investigate varying the number of generations between dispersal episodes, $T$. We first consider the case in which anti-social punishment is removed, giving initial starting frequencies of 25\% pro-social punishers, 25\% non-punishing cooperators, and 50\% non-punishing defectors. Figure~\ref{figTvsGSnoASP} shows the effect of $T$ on the largest group size for which the equilibrium is polymorphic for pro-social punishment and non-punishing cooperation, i.e. where defection is removed by selection. We took a simulation run to reach this equilibrium when the global proportion of pro-social punishers and cooperators exceeded 75\% after 1000 cycles. The shading in the figure indicates the number of simulation runs, out of 100, which reached this equilibrium.

The results show that the largest group size for which pro-social punishment and cooperation are selected increases with $T$. At first sight this result is surprising, since in classic Haystack models without punishment a large $T$ eliminates cooperation \citep{Wilson:1987:a,Sober:1998:a}. This is because in such models there is only a single equilibrium within-groups: all defect. Thus, in the limit of $T$ approaching infinity any group founded by at least one defector will be converted to all defect. Frequent dispersal can, though, keep the population out of this equilibrium and allow cooperation to be stable globally \citep{Sober:1998:a}. However, the addition of pro-social punishment to such models means that groups founded by one or more defectors need not, for high $T$, be converted to all defect. 

For illustration, consider groups consisting of pro-social punishers
and non-punishing defectors. Pro-social punishment is then a stable
equilibrium within the group whenever the proportion of punishers
causes the condition $Px_p > Kx_d + C$ to be satisfied. Since a single
group constitutes a well-mixed population, i.e. each group member interacts with all other group members with respect to public goods, this occurs above the limit of $\frac{C+K}{P+K}$ (see analytical results). Under the parameters used in the simulations, this is when the fraction of pro-social punishers when the group is founded is greater than $1/3$. Groups founded by more than this proportion of pro-social punishers will thus be stable against defection. Thus, a high $T$ need not cause an increase in defection within each group. Moreover, individuals in groups at the pro-social punishment equilibrium have a higher mean fitness, due to the benefits of cooperation (Equations~\ref{eqnWd}--\ref{eqnWp}). The more generations the group stays together for, the greater the cumulative total of this benefit compared to non-cooperation groups, since groups grow at an exponential rate (with no negative density-dependent effects) in accordance with Equation~\ref{eqnReproduction}. Consequently, when dispersal eventually occurs groups at the pro-social punishment equilibrium will have grown to a larger size, and will hence make up a larger fraction of the migrant pool. 

A larger $T$ can thus favour pro-social punishment, due to the
reproductive advantage that individuals in groups at the pro-social
punishment equilibrium enjoy with each generation that the group stays
together. This mechanism depends upon some groups being founded with
type frequencies that fall within the basin of attraction for the
pro-social punishment equilibrium, i.e. with $x_p>1/3$. This can
occur in a group-structured population even when the global value of
$x_p$ is below this (i.e.~0.25 as used in the simulations), provided
there is variance in the composition of groups when they are
founded. In our model, this variance is provided through the formation
of groups by hypergeometric (random) sampling of individuals from the
migrant pool. This variance decreases, however, as the founding size
of the groups becomes larger. Thus for larger group sizes, fewer
groups lie in the basin of attraction for the pro-social punishment
equilibrium. This explains why for
large group sizes, a large $T$ may not be sufficient to select for
pro-social punishment, because too few (or no) groups may fall in its
basin of attraction.  %
Finally, we consider the effects of varying $T$ as well as group size
when anti-social punishment is added to the model. In
Figure~\ref{figTvsGSwithASP} we use the same parameters as in
Figure~\ref{figTvsGSnoASP}, but here we reintroduce anti-social
punishment, returning to the initial condition in which all four
strategies are at 25\%. %
In this case we see that pro-social punishment and cooperation are
selected over a much smaller range of group sizes for a given
dispersal frequency. As $T$ is increased, the increase in the largest
group size under which pro-social punishment and cooperation are
selected is also less pronounced.

The reason for this result is that the basin boundary for the
pro-social punishment equilibrium changes within a group. In
particular, from the analytical results we know that in a single group
founded by pro-social punishers and anti-social punishers only, the
basin boundary lies at $x_p=1-\frac{C+K+P}{2\left(P+K\right)}$, which
is a more stringent condition than the population without anti-social
punishment ($\frac{C+K}{P+K}$). Under the parameters used in the
simulation, this corresponds to $x_p>1/3$ without anti-social
punishment, and $x_p>5/12$ in the presence of anti-social
punishment. We verified numerically that this result holds in the
four-type system when the three other types are all started in equal
proportion. Thus for a given frequency of pro-social punishment in the
migrant pool, fewer groups are expected to fall in the basin of
attraction for the pro-social punishment equilibrium when some of the
non-punishing defectors are replaced with anti-social punishers. The
presence of anti-social punishment therefore means that a larger
between-group variance (relative to the migrant pool value of $x_p$)
is required to select for pro-social punishment.%
\subsection{Sensitivity to the effect-to-cost ratio of cooperation
  and punishment}
\label{sensitivitySec}
The benefit-to-cost ratio of the cooperative act affects the largest
group size in which cooperation can evolve. Without either type of
punishment, when $T=1$ then cooperation is selectively advantageous
when $B/n>C$; this is the well-known result from
Wilson's \citeyearpar{Wilson:1975:a} trait-group model when groups are
formed randomly. When both pro- and anti-social punishment are
available, and all four strategies are initially present in the
population at equal frequency, then our simulations shows that the
effects of both type of punishment on the level of cooperation
cancel out. In other words, the condition for cooperation to be
favoured is still $B/n>C$ when both types of punishment are added and
$T=1$. We find this result to be insensitive to the effect-to-cost
ratio of punishment; it holds in the simulations even for $P=K=0.1$,
as well as for $P>K$.

For $T>1$, we find that the addition of anti-social punishment
reduces the largest group size over which cooperation evolves,
compared to the case where only pro-social punishment was available.
This result is also qualitatively insensitive to the benefit-to-cost
ratio of cooperation. That is, decreasing the benefit-to-cost ratio of
cooperation decreases the largest group size in which cooperation
evolves in the case where both anti- and pro-social punishment are
present, and similarly in the case where only pro-social punishment is
available. The result that anti-social punishment further reduces this
over the pro-social punishment only case, however, still holds
regardless of the $B/C$ ratio when $K=0.1$ and $P=0.5$. Further, we
found that increasing $P$ (while holding $K$ constant) increases the
magnitude of this effect. That is, the addition of anti-social
punishment makes a greater difference as $P$ increases. As $P$
decreases, however, both types of punishment have less effect and
the difference between the two cases becomes smaller. When $K=0.1$ and
$P<0.4$, we found that the addition of anti-social punishment made no
difference to the range of group sizes over which cooperation evolves,
compared to the case where only pro-social punishment is available.

It should be noted that $P>K$ is a common assumption in both models of
the evolution of punishment
(e.g. \citealt{Boyd:2003:a,Boyd:2010:a,Bowles:2004:a,Lehmann:2007:c,dosSantos:2011:a}),
and in experimental public goods games
\citep{Fehr:2002:a,Herrmann:2008:a}. Nevertheless, measuring the
actual cost-to-effect ratio of punishment \textit{in situ} in real
populations is very difficult. Indeed, some authors have explicitly
considered the case where $P=K$ \citep{Rand:2010:a}. However, at least
in the case of humans it is commonly held that the advent of tools
from gossip to weaponry makes punishment very effective at little cost
to the punisher \citep{Sober:1998:a,Bingham:1999:a,Boehm:1999:a,Binmore:2005:a}. We have thus
focused our study on cases where punishment is reasonably efficient in
terms of effect-to-cost ratio, whilst still being less efficient than
the benefit-to-cost ratio of the cooperative act.

\section{Discussion and conclusion}
\label{disc.sec}

We have presented here, to our knowledge, the first model of the
evolution of anti-social punishment in group-structured
populations. Previous work on anti-social punishment has nevertheless
suggested that group structure would favour pro-social punishment and
prevent anti-social punishment from being effective
\citep{Rand:2010:a,Rand:2011:a}. Further, much previous work on the
evolution of pro-social punishment in group-structured populations
has not even considered the possibility of anti-social punishment
(e.g. \citealt{Boyd:2003:a,Boyd:2010:a,Gardner:2004:a,Lehmann:2007:c}),
presumably for the same reason. After all, group or kin selection should be expected to promote behaviours that support
cooperation rather than defection, assuming
a population structure that provides positive relatedness at the locus
for cooperation \citep{Hamilton:1964:a,Wilson:1975:a}.

In fact, we have shown here that anti-social punishment \emph{can}
be effective in the preventing the evolution of pro-social punishment
and cooperation in group-structured populations. Models for the
evolution of pro-social punishment typically rely both on pro-social
punishment being a stable equilibrium within a group, and on some
groups being founded with initial strategy frequencies that fall
within the basin of attraction for this equilibrium. When these two
conditions are met, equilibrium selection between groups can occur
\citep{Harsanyi:1988:a,Boyd:1990:a,Binmore:1998:a,Canals:1998:a}, such that groups at the
pro-social punishment and cooperation equilibrium out-compete those at
the defection equilibrium \citep{Boyd:2003:a}. We have shown here that
the presence of anti-social punishers reduces the likelihood of the
second condition being met, by reducing the basin of attraction for
the pro-social punishment equilibrium, compared to a population where
the anti-social punishers are replaced with non-punishing
defectors. Consequently, a greater between-group variance in the
frequency of pro-social punishment is required in order for some
groups to fall in its basin of attraction, and equilibrium selection
to occur. 

One way such a greater between-group variance can be achieved is
through a reduction in group size. However,  such a
requirement eliminates punishment as an explanation for the
maintenance of cooperation in large human groups with low relatedness.
Notice though that the results in this model assume that there is no
structure to a society {\em within} a large group.  However,
internal structure within large groups may in fact be present e.g. due to social hierarchy or
spatial distribution of public goods.  Nevertheless, the homogeneity of
within-group structure is a standard assumption in group-structured models of the evolution of cooperation \citep{Boyd:1990:a,Boyd:2003:a,Boyd:2010:a,Wilson:1975:a,Wilson:1987:a,Hamilton:1975:a,Traulsen:2006:a,Lehmann:2007:c}, and corresponds to a public good that is shared equally with all group members. Moreover, if the effective group size of social interactions is smaller then pro-social punishment may not be necessary to maintain cooperation anyway; direct and indirect fitness benefits from the cooperative act itself may be sufficient. We have thus focused here on cases where the public good is shared equally between all group members and hence pro-social punishment \emph{is} necessary to maintain cooperation in large groups. For the same reason, we have also focused on linear public goods games. This is because in non-linear public goods games, punishment would not be needed to maintain cooperation even in large randomly formed groups \citep{Archetti:2012:a}. Whether real-life social dilemmas are linear or non-linear is an empirical question that must be answered on a case by case basis.

It is worth stressing that we have considered the \emph{maintenance}
of punishment, rather than its invasion from mutation frequency. It is
already widely appreciated that punishment can be stable when common,
even if it is not selected when rare
\citep{Boyd:1992:a,Lehmann:2007:c}. This is because in the standard
model, %
the total individual cost of punishment decreases as punishers
increase in frequency within a group
(Equations~\ref{eqnWc}--\ref{eqnWa}). Thus, both pro- and anti-social
punishment undergo positive frequency-dependent selection. In light of
this, much work has focused on the maintenance of punishment (e.g.
\citealt{Boyd:1992:a,Henrich:2001:a,Gintis:2003:a}), as we do in this
study. Several mechanisms have, however, been suggested for the
invasion of (pro-social) punishment from rarity. These include kin
benefits resulting from punishing acts reducing local competition
\citep{Lehmann:2007:c}, the fixation of punishment within a single
group through stochastic processes \citep{Boyd:2003:a}, voluntary
participation in social interactions \citep{Hauert:2007:a}, systems of
reputation \citep{dosSantos:2011:a}, or the coordination of punishing
behaviour between individuals
\citep{Boyd:2010:a}. It has been shown more
generally that social traits which are maladaptive when rare, but
advantageous once common, may be able to reach the threshold frequency
for positive selection by drift-like processes
\citep{Boyd:1990:a,Boyd:2003:a}. One mechanism by which this may occur
is when environmental factors result in population oscillations, and
periods where the environment is temporarily below its maximum
carrying capacity \citep{Cace07,Alizon08}. %
Similarly, future studies should investigate the proximate mechanisms
by which anti-social punishment might be favoured over simple
non-punishing defection within a single group. This is similar to
the %
classic problem of how pro-social punishers may be favoured over
non-punishing cooperators within a single group
\citep{Colman:2006:a}. Essentially, either type of punishment can be
favoured if the effects of punishment are not shared equally with
non-punishers in the same group, but instead feed disproportionately
back to the actor or their
kin. %
As mentioned earlier, it is difficult to imagine direct advantage from
anti-social punishment, at least as described here and in the
economics literature.  However, a linked consequence such as increased
social status may serve as the explanatory benefit.

In conclusion, we have shown here that the presence of anti-social
punishers reduces the range of conditions over which pro-social
punishment and cooperation are stable in group-structured
populations. This occurs because anti-social punishment reduces the
basin of attraction for the pro-social punishment equilibrium within
groups. Thus, a given magnitude of between-group variance may no
longer be sufficient to select for pro-social punishment. In
particular, we have shown here how the range of group sizes over which
pro-social punishment is selected can be greatly reduced by
anti-social punishment. Given the existence of anti-social punishment
in all studied extant human cultures \citep{Herrmann:2008:a,SylwesterHomo}, our results suggest that the
claims of models showing the evolution of pro-social punishment in
group-structured populations should be re-evaluated with the addition
of anti-social punishment.

\section*{Acknowledgments}
We thank Laurent Lehmann, and members of the AmonI group at the
University of Bath --- particularly Karolina Sylwester --- for useful discussions. We also thank Marco Archetti and an anonymous reviewer for helpful feedback, and Slimane Dridi and Jorge Pe\~{n}a for comments on a draft of the manuscript.
This effort was sponsored by the US Air Force Office of Scientific
Research, Air Force Material Command, USAF, under grant number
FA8655-10-1-3050.

\bibliographystyle{elsarticle-harv}
\bibliography{lit}

\begin{thebibliography}{60}
\expandafter\ifx\csname natexlab\endcsname\relax\def\natexlab#1{#1}\fi
\expandafter\ifx\csname url\endcsname\relax
  \def\url#1{\texttt{#1}}\fi
\expandafter\ifx\csname urlprefix\endcsname\relax\def\urlprefix{URL }\fi

\bibitem[{Alizon and Taylor(2008)}]{Alizon08}
Alizon, S., Taylor, P., 2008. Empty sites can promote altruistic behaviour.
  Evolution 62~(6), 1335--1344.

\bibitem[{Anderson and Putterman(2006)}]{AndersonPutterman06}
Anderson, C.~M., Putterman, L., 2006. {Do non-strategic sanctions obey the law
  of demand? The demand for punishment in the voluntary contribution
  mechanism}. Game. Econ. Behav. 54~(1), 1--24.

\bibitem[{Archetti and Scheuring(2012)}]{Archetti:2012:a}
Archetti, M., Scheuring, I., 2012. Review: Game theory of public goods in
  one-shot social dilemmas without assortment. J. Theor. Biol. 299, 9--20.

\bibitem[{Bergstrom(2002)}]{Bergstrom:2002:a}
Bergstrom, T.~C., 2002. Evolution of social behavior: Individual and group
  selection. J. Econ. Perspect. 16~(2), 67--88.

\bibitem[{Bingham(1999)}]{Bingham:1999:a}
Bingham, P., 1999. Human uniqueness: a general theory. Q. Rev. Biol., 133--169.

\bibitem[{Binmore(1998)}]{Binmore:1998:a}
Binmore, K., 1998. Just Playing: Game Theory and the Social Contract. Vol.~2.
  MIT press.

\bibitem[{Binmore(2005)}]{Binmore:2005:a}
Binmore, K., 2005. Natural Justice. Oxford University Press.

\bibitem[{Boehm(1999)}]{Boehm:1999:a}
Boehm, C., 1999. Hierarchy in the Forest: The Evolution of Egalitarian
  Behavior. Harvard Univ Pr.

\bibitem[{Bowles and Gintis(2004)}]{Bowles:2004:a}
Bowles, S., Gintis, H., 2004. The evolution of strong reciprocity: cooperation
  in heterogeneous populations. Theor. Popul. Biol. 65~(1), 17--28.

\bibitem[{Boyd et~al.(2010)Boyd, Gintis, and Bowles}]{Boyd:2010:a}
Boyd, R., Gintis, H., Bowles, S., 2010. Coordinated punishment of defectors
  sustains cooperation and can proliferate when rare. Science 328~(5978), 617.

\bibitem[{Boyd et~al.(2003)Boyd, Gintis, Bowles, and Richerson}]{Boyd:2003:a}
Boyd, R., Gintis, H., Bowles, S., Richerson, P.~J., 2003. The evolution of
  altruistic punishment. Proc. Natl. Acad. Sci. U. S. A. 100~(6), 3531--3535.

\bibitem[{Boyd and Richerson(1992)}]{Boyd:1992:a}
Boyd, R., Richerson, P., 1992. Punishment allows the evolution of cooperation
  (or anything else) in sizable groups. Ethol. Sociobiol. 13~(3), 171--195.

\bibitem[{Boyd and Richerson(1990)}]{Boyd:1990:a}
Boyd, R., Richerson, P.~J., 1990. Group selection among alternative
  evolutionarily stable strategies. J. Theor. Biol. 145, 331--342.

\bibitem[{Canals and Vega-Redondo(1998)}]{Canals:1998:a}
Canals, J., Vega-Redondo, F., 1998. Multi-level evolution in population games.
  International Journal of Game Theory 27~(1), 21--35.

\bibitem[{Cohen et~al.(1976)Cohen, Eshel, et~al.}]{Cohen:1976:a}
Cohen, D., Eshel, I., et~al., 1976. On the founder effect and the evolution of
  altruistic traits. Theor. Popul. Biol. 10~(3), 276.

\bibitem[{Colman(2006)}]{Colman:2006:a}
Colman, A., 2006. The puzzle of cooperation. Nature 440~(7085), 744--745.

\bibitem[{dos Santos et~al.(2011)dos Santos, Rankin, and
  Wedekind}]{dosSantos:2011:a}
dos Santos, M., Rankin, D., Wedekind, C., 2011. The evolution of punishment
  through reputation. Proc. R. Soc. Biol. Sci. Ser. B 278~(1704), 371--377.

\bibitem[{Eldakar and Wilson(2008)}]{Eldakar08}
Eldakar, O.~T., Wilson, D.~S., 2008. Selfishness as second-order altruism.
  Proceedings of the National Academy of Sciences 105~(19), 6982--6986.

\bibitem[{Fehr et~al.(2002)Fehr, Fischbacher, and G{\"a}chter}]{Fehr:2002:a}
Fehr, E., Fischbacher, U., G{\"a}chter, S., 2002. Strong reciprocity, human
  cooperation, and the enforcement of social norms. Hum. Nature-int. Bios.
  13~(1), 1--25.

\bibitem[{Fletcher and Zwick(2004)}]{Fletcher:2004:a}
Fletcher, J.~A., Zwick, M., 2004. Strong altruism can evolve in randomly formed
  groups. J. Theor. Biol. 228, 303--313.

\bibitem[{Fletcher and Zwick(2007)}]{Fletcher:2007:a}
Fletcher, J.~A., Zwick, M., 2007. The evolution of altruism: Game theory in
  multilevel selection and inclusive fitness. J. Theor. Biol. 245~(1), 26--36.

\bibitem[{Foster et~al.(2006)Foster, Wenseleers, and Ratnieks}]{Foster:2006:a}
Foster, K.~R., Wenseleers, T., Ratnieks, F.~L., 2006. Kin selection is the key
  to altruism. Trends Ecol. Evol. 21~(2), 57--60.

\bibitem[{Frank(1998)}]{Frank:1998:a}
Frank, S.~A., 1998. Foundations of Social Evolution. Monographs in Behavior and
  Ecology. Princeton University Press, Princeton.

\bibitem[{Gardner and West(2004)}]{Gardner:2004:a}
Gardner, A., West, S., 2004. Cooperation and punishment, especially in humans.
  Am. Nat. 164~(6), 753--764.

\bibitem[{Gintis et~al.(2003)Gintis, Bowles, Boyd, and Fehr}]{Gintis:2003:a}
Gintis, H., Bowles, S., Boyd, R., Fehr, E., 2003. Explaining altruistic
  behavior in humans. Evol. Hum. Behav. 24~(3), 153--172.

\bibitem[{Griffin et~al.(2004)Griffin, West, and Buckling}]{Griffin:2004:a}
Griffin, A.~S., West, S.~A., Buckling, A., 2004. Cooperation and competition in
  pathogenic bacteria. Nature 430, 1024--1027.

\bibitem[{Hamilton(1964)}]{Hamilton:1964:a}
Hamilton, W.~D., 1964. {The genetical evolution of social behaviour. I}. J.
  Theor. Biol. 7~(1), 1--16.

\bibitem[{Hamilton(1975)}]{Hamilton:1975:a}
Hamilton, W.~D., 1975. Innate social aptitudes in man, an approach from
  evolutionary genetics. In: Fox, R. (Ed.), Biosocial Anthropology. Malaby
  Press, pp. 133--155.

\bibitem[{Hammerstein(2003)}]{Hammerstein:2003:a}
Hammerstein, P., 2003. Genetic and Cultural Evolution of Cooperation. MIT
  Press.

\bibitem[{Hardin(1968)}]{Hardin:1968:a}
Hardin, G., 1968. The tragedy of the commons. Science 162, 1243--1248.

\bibitem[{Harsanyi and Selten(1988)}]{Harsanyi:1988:a}
Harsanyi, J.~C., Selten, R., 1988. A General Theory of Equilibrium Selection in
  Games. MIT Press, Cambridge, MA.

\bibitem[{Hauert et~al.(2007)Hauert, Traulsen, Brandt, Nowak, and
  Sigmund}]{Hauert:2007:a}
Hauert, C., Traulsen, A., Brandt, H., Nowak, M., Sigmund, K., 2007. Via freedom
  to coercion: the emergence of costly punishment. Science 316~(5833),
  1905--1907.

\bibitem[{Henrich and Boyd(2001)}]{Henrich:2001:a}
Henrich, J., Boyd, R., 2001. Why people punish defectors:: Weak conformist
  transmission can stabilize costly enforcement of norms in cooperative
  dilemmas. J. Theor. Biol. 208~(1), 79--89.

\bibitem[{Herrmann et~al.(2008)Herrmann, Th{\"o}ni, and
  G{\"a}chter}]{Herrmann:2008:a}
Herrmann, B., Th{\"o}ni, C., G{\"a}chter, S., 2008. Antisocial punishment
  across societies. Science 319~(5868), 1362.

\bibitem[{Killingback et~al.(2006)Killingback, Bieri, and
  Flatt}]{Killingback:2006:a}
Killingback, T., Bieri, J., Flatt, T., 2006. Evolution in group-structured
  populations can resolve the tragedy of the commons. Proc. R. Soc. Biol. Sci.
  Ser. B 273, 1477--1481.

\bibitem[{Kreft(2004)}]{Kreft:2004:a}
Kreft, J.-U., 2004. Biofilms promote altruism. Microbiology 150, 2751--2760.

\bibitem[{Lehmann and Keller(2006)}]{Lehmann:2006:a}
Lehmann, L., Keller, L., 2006. The evolution of cooperation and altruism - a
  general framework and a classification of models. J. Evol. Biol. 19~(5),
  1365--1376.

\bibitem[{Lehmann et~al.(2007)Lehmann, Rousset, Roze, and
  Keller}]{Lehmann:2007:c}
Lehmann, L., Rousset, F., Roze, D., Keller, L., 2007. {Strong reciprocity or
  strong ferocity? A population genetic view of the evolution of altruistic
  punishment}. Am. Nat. 170~(1), 21--36.

\bibitem[{Mathew and Boyd(2011)}]{Mathew:2011:a}
Mathew, S., Boyd, R., 2011. Punishment sustains large-scale cooperation in
  prestate warfare. Proc. Natl. Acad. Sci. U. S. A. 108~(28), 11375.

\bibitem[{{Maynard Smith}(1964)}]{Smith:1964:a}
{Maynard Smith}, J., 1964. Group selection and kin selection. Nature 201,
  1145--1147.

\bibitem[{{Maynard Smith} and Szathm\'{a}ry(1995)}]{Smith:1995:a}
{Maynard Smith}, J., Szathm\'{a}ry, E., 1995. Major Transitions in Evolution.
  W. H. Freeman/Spektrum, Oxford.

\bibitem[{Michod(1983)}]{Michod:1983:a}
Michod, R.~E., 1983. Evolution of the first replicators. American Zoology 23,
  5--14.

\bibitem[{Nunney(1985)}]{Nunney:1985:a}
Nunney, L., 1985. Group selection, altruism, and structured-deme models. Am.
  Nat. 126~(2), 212--230.

\bibitem[{Okasha(2006)}]{Okasha:2006:a}
Okasha, S., 2006. Evolution and the Levels of Selection. Clarendon Press.

\bibitem[{Pepper(2000)}]{Pepper:2000:a}
Pepper, J.~W., 2000. Relatedness in trait group models of social evolution. J.
  Theor. Biol. 206~(3), 355--368.

\bibitem[{Pfeiffer et~al.(2001)Pfeiffer, Schuster, and
  Bonhoeffer}]{Pfeiffer:2001:a}
Pfeiffer, T., Schuster, S., Bonhoeffer, S., 2001. Cooperation and competition
  in the evolution of {ATP}-producing pathways. Science 292~(5516), 504--507.

\bibitem[{Powers et~al.(2011)Powers, Penn, and Watson}]{Powers:2011:a}
Powers, S.~T., Penn, A.~S., Watson, R.~A., 2011. The concurrent evolution of
  cooperation and the population structures that support it. Evolution 65~(6),
  1527–1543.

\bibitem[{Rand et~al.(2010)Rand, Armao~IV, Nakamaru, and Ohtsuki}]{Rand:2010:a}
Rand, D., Armao~IV, J., Nakamaru, M., Ohtsuki, H., 2010. Anti-social punishment
  can prevent the co-evolution of punishment and cooperation. J. Theor. Biol.
  265~(4), 624--632.

\bibitem[{Rand and Nowak(2011)}]{Rand:2011:a}
Rand, D., Nowak, M., 2011. The evolution of antisocial punishment in optional
  public goods games. Nature Communications 2, 434.

\bibitem[{Santos and Szathm\'{a}ry(2008)}]{Santos:2008:a}
Santos, M., Szathm\'{a}ry, E., 2008. Genetic hitchhiking can promote the
  initial spread of strong altruism. BMC Evol. Biol. 8~(281).

\bibitem[{Sober and Wilson(1998)}]{Sober:1998:a}
Sober, E., Wilson, D.~S., 1998. Unto Others: The Evolution and Psychology of
  Unselfish Behavior. Harvard University Press, Cambridge, MA.

\bibitem[{Sylwester et~al.(2011)Sylwester, Herrmann, and
  Bryson}]{SylwesterHomo}
Sylwester, K., Herrmann, B., Bryson, J.~J., 2011. {\em Homo homini lupus?} an
  evolutionary view on antisocial punishment., under review.

\bibitem[{Szathm\'{a}ry(2011)}]{Szathmary:2011:a}
Szathm\'{a}ry, E., 2011. To group or not to group? Science 324~(19),
  1648--1649.

\bibitem[{Traulsen and Nowak(2006)}]{Traulsen:2006:a}
Traulsen, A., Nowak, M., 2006. Evolution of cooperation by multilevel
  selection. Proc. Natl. Acad. Sci. U. S. A. 103~(29), 10952--10955.

\bibitem[{\v{C}a\v{c}e and Bryson(2007)}]{Cace07}
\v{C}a\v{c}e, I., Bryson, J., 2007. Agent based modelling of communication
  costs: {W}hy information can be free. In: Lyon, C., Nehaniv, C.~L.,
  Cangelosi, A. (Eds.), Emergence and Evolution of Linguistic Communication.
  Springer, London, pp. 305--322.

\bibitem[{West et~al.(2007)West, Griffin, and Gardner}]{West:2007:b}
West, S., Griffin, A., Gardner, A., 2007. Social semantics: altruism,
  cooperation, mutualism, strong reciprocity and group selection. J. Evol.
  Biol. 20, 415--432.

\bibitem[{Wilson(1975)}]{Wilson:1975:a}
Wilson, D.~S., 1975. A theory of group selection. Proc. Natl. Acad. Sci. U. S.
  A. 72~(1), 143--146.

\bibitem[{Wilson(1979)}]{Wilson:1979:a}
Wilson, D.~S., 1979. Structured demes and trait-group variation. Am. Nat.
  113~(4), 606--610.

\bibitem[{Wilson(1987)}]{Wilson:1987:a}
Wilson, D.~S., 1987. Altruism in mendelian populations derived from sibling
  groups: The {Haystack} model revisited. Evolution 41~(5), 1059--1070.

\bibitem[{Wilson(1990)}]{Wilson:1990:a}
Wilson, D.~S., 1990. Weak altruism, strong group selection. Oikos 59~(1),
  135--140.

\end{thebibliography}

\end{document}